# Title

- Observation of Kondo hybridization with an orbital-selective Mott phase in $4d$ $Ca_{2-x}Sr_xRuO_4$

# Authors

Minsoo Kim,[1,2] Junyoung Kwon,[1,2] Choong H. Kim,[1,2] Younsik Kim,[1,2] Daun Chung,[3] Hanyoung Ryu,[1,2] Jongkeun Jung,[1,2] Beom Seo Kim,[1,2] Dongjoon Song,[1,2] Jonathan D. Denlinger,[4] Moonsup Han,[5] Yoshiyuki Yoshida,[6] Takashi Mizokawa,[7] Wonshik Kyung,[1,2]* Changyoung Kim[1,2]*

# Affiliations

[1]Center for Correlated Electron Systems, Institute for Basic Science (IBS), Seoul 08826, Republic of Korea.
[2]Department of Physics & Astronomy, Seoul National University, Seoul 08826, Republic of Korea.
[3]College of Liberal Studies, Seoul National University, Seoul 08826, Republic of Korea.
[4]Advanced Light Source (ALS), Lawrence Berkeley National Laboratory, Berkeley, California 94720, USA.
[5]Department of Physics, University of Seoul, Seoul 02504, Republic of Korea.
[6]National Institute of Advanced Industrial Science and Technology (AIST), Tsukuba 305-8568, Japan.
[7]Department of Applied Physics, Waseda University, Tokyo 169-8555, Japan.

*These authors contributed equally to this work.
*Corresponding author. Email: specialtoss@gmail.com (W.K.); changyoung@snu.ac.kr (C.K.)

# Abstract

The heavy fermion state with Kondo-hybridization (KH), usually manifested in $f$-electron systems with lanthanide or actinide elements, was recently discovered in several $3d$ transition metal compounds without $f$-electrons. However, KH has not yet been observed in $4d/5d$ transition metal compounds, since more extended $4d/5d$ orbitals do not usually form flat bands that supply localized electrons appropriate for Kondo pairing. Here, we report a doping- and temperature-dependent angle-resolved photoemission study on $4d$ $Ca_{2-x}Sr_xRuO_4$, which shows the signature of KH. We observed a spectral weight transfer in the γ-band, reminiscent of an orbital-selective Mott phase (OSMP). The Mott localized γ-band induces KH with the itinerant β-band, resulting in spectral weight suppression around the Fermi level. Our work is the first to demonstrate the evolution of the OSMP with possible KH among $4d$ electrons, and thereby expands the material boundary of Kondo physics to $4d$ multi-orbital systems.

# MAIN TEXT

## Introduction

The heavy fermion (HF) state is one of the most important subjects in strongly correlated systems research and is often accompanied by exotic states such as superconductivity, quantum criticality and magnetism (*1*). In early studies, HF behavior was mostly found in *f*-electron systems in which strongly localized *f*- and itinerant *spd*- hybridized orbitals coexist and pair to form Kondo-singlets (*2*). Recently, unexpected HF states have been discovered in moderately localized 3*d*-electron systems such as $CaCu_3Ir_4O_{12}$, $AFe_2As_2$ (A = K, Rb, Cs), and $Fe_3GeTe_2$ (*3-5*), providing a strong impetus to search for possible Kondo-pairing even in less localized 4*d*/5*d*-electron systems.

$Ca_{2-x}Sr_xRuO_4$ (CSRO), a 4*d* transition metal oxide (TMO), is a good candidate for a HF state. Previous experimental studies (*6-8*) showed that CSRO possesses HF-like quasiparticle states, which suggests the existence of an orbital-selective Mott phase (OSMP). This led to several angle-resolved photo-emission spectroscopy (ARPES) studies that aimed to obtain direct evidence of the OSMP in CSRO (*9-11*), but controversy still remains even regarding the very existence of the OSMP. Part of the reason for the inconsistent conclusions arises from the fact that those studies were performed with limited points in the parameter space, such as doping concentration ($x$) and temperature ($T$).

In this article, we provide our systematic $x$- ($0.2 \leq x \leq 0.5$) and $T$-dependent ARPES results on CSRO. With variations in $x$ and $T$, a gradual orbital-selective opening of a soft gap is observed in the γ ($4d_{xy}$)-band with spectral weight transfer from lower- to higher-binding energy (BE), suggesting the emergence of the OSMP. We also observed unexpected spectral weight suppression of the β ($4d_{xz/yz}$)-band (but not α), which we attribute to Kondo-hybridization (KH) between the localized γ- and itinerant β-bands, based on previous studies (*4-8, 12-16*). Our results show the coincidence between the emergence of the OSMP (and KH) and octahedral tilting distortion, implying that tilting is the key parameter triggering the OSMP, as well as KH. Our results not only provide direct evidence of the OSMP, but also constitute the first demonstration of possible KH in 4*d*-orbitals.

## Results

### Experimental evidence for orbital-selective Mott phase

CSRO takes two types of $RuO_2$ octahedral distortions: octahedral rotation (in-plane rotation about the c-axis) and tilting (out-of-plane polar rotation about the b-axis). Thus, there are three crystalline forms in terms of the distortions (*17*) (see section S2): (I) neither rotation nor tilting ($1.5 \leq x \leq 2.0$, *I4/mmm*), (II) finite rotation without tilting ($0.5 \leq x < 1.5$, *I4$_1$/acd*), and (III) finite rotation and tilting ($0 \leq x < 0.5$, *Pbca*). The Fermi surface (FS) topology varies significantly depending on the type of distortion (Fig. 1). In $Sr_2RuO_4$ (I), four electrons ($4d^4$) in the $t_{2g}$ orbitals make up three FS pockets (Fig. 1A). When octahedral rotation occurs (II), the FS becomes zone-folded due to the reduced BZ (Fig.

1B). Finally, in III with octahedral tilting, the β- and γ-FS pockets are selectively suppressed while the α-pocket remains robust as seen in Fig. 1C. This behavior is reminiscent of the OSMP.

To scrutinize this OSMP-like phenomenon, we performed systematic ARPES studies as a function of $x$ and $T$. Our $x$-dependent result ($0.2 \leq x \leq 0.5$) is presented in Fig. 2 (A-D). As $x$ decreases from $x = 0.5$ to 0.2, gradual suppression of the spectral weight is observed near the Fermi level generating a soft gap (*18*), as shown in Fig. S2B (see section S2). Interestingly, the soft gap opens only for the β- and γ-bands, while the α-band remains intact with variation in doping. A similar trend is also observed in the $T$-dependence data (Fig. 2 (D-H)). As $T$ decreases from 45 K, the spectral weight of the β- and γ-bands is suppressed in a similar fashion to that observed for the doping dependent result.

A detailed analysis of the spectral weight suppression in Fig. 2 is provided in Fig. 3. It can be clearly seen in the momentum distribution curves (MDCs) for each $x$ (Fig. 3A) that the β- and γ-bands are selectively suppressed as a function of $x$. The Lorentzian-fitted peak areas are plotted in Figs. 3B ($x$-dependent) and C ($T$-dependent), showing clear suppression only for the β- and γ-bands. The energy distribution curves (EDCs) in Fig. 3D also show the evolution of the soft gap in the γ-band as a function of $T$. As $T$ decreases from 45 K, the spectral weight in the 'A' region (BE < 0.2 eV) is gradually suppressed, while it increases in the 'B' region (BE ~ 0.4 eV). Hence, the spectral weight is transferred from a lower to higher BE region. This behavior can be seen more clearly in Fig. 3E, which shows the EDCs with the data obtained at 45 K subtracted. The integrated areas in the 'A' and 'B' regions are found to be almost identical, but with the opposite sign, satisfying the sum rule at all $T$ (Fig. 3F). This suggests that the $T$-dependent evolution originates from (Mott-like) spectral weight transfer rather than a $T$-driven spectral broadening effect (*19*).

To confirm the origin of the spectral weight suppression further, we investigated the electronic structure change for $x = 0.2$ upon surface electron doping (Fig. 4C and section S6). In previous studies (*20-22*), it has been revealed that the Mott insulating state can easily collapse with an infinitesimal electron doping, showing emergence of a quasiparticle peak. Consistent with the results, we observe appearance of clear quasiparticle bands upon doping of small amount of electron (Fig. 4C). Therefore, we conclude that the spectral weight suppression indeed originates from Mott localization (*20*).

**Role of octahedral tilting and signature of possible Kondo-hybridization**

Our observations (orbital-selective suppression and spectral weight transfer) are consistent with the previously proposed OSMP scenario (*23-25*). However, there are still questions to be answered, e.g., what triggers OSMP and why the β-band is suppressed in the same way as the γ-band. It was suggested in previous experimental and theoretical studies that octahedral rotation is responsible for the OSMP (*10, 23*). A study suggested that the rotation sufficiently reduces the γ-band width for Mott localization (*23*) while the importance of $\sqrt{2} \times \sqrt{2} \times 2$ unit cell doubling due to the rotation was discussed in another (*10*). In both scenarios, octahedral rotation plays a significant role in the OSMP.

Therefore, a larger rotation angle may lead to a stronger OSMP effect. However, the OSMP occurs after the octahedral rotation angle saturates to the maximum value at $x = 0.5$. Our results show that the OSMP and octahedral tilting distortion appear coincidently, and that the strength of the OSMP (spectral weight suppression of the γ-band) is roughly proportional to the tilting angle (*17, 23*). Therefore, even though octahedral rotation significantly reduces the bandwidth of the γ-band, the OSMP in CSRO is triggered by the octahedral tilting, not by the rotation.

The mechanism with which octahedral tilting triggers the OSMP can be understood by considering the effect of the octahedral distortions (rotation/tilting) to bandwidths (Fig. 5). The key aspect of octahedral distortions is that they lead to narrower *d*-orbital bandwidths. The detailed explanation for the band width reduction with octahedral distortions is as follows. Without octahedral rotation/tilting, 4*d* orbitals (three $t_{2g}$ and two $e_g$) of $Sr_2RuO_4$ possess a wide bandwidth (Fig. 5 (A-C)). Once octahedral (in-plane) rotation sets in, the $d_{xy}$ and $d_{x2-y2}$ orbitals become hybridized, which leads to a bandwidth reduction in the γ ($d^*_{xy}$) band (*23*) while the $d_{yz/xz}$ orbitals remain almost unchanged (Fig. 5 (D-F)). On top of that, octahedral tilting leads to hybridization between $d_{xy}$ (γ-band) and $d_{yz/zx}$ orbitals (α-, β-bands). The hybridization from the tilt distortion results in fragmented bands with narrower bandwidths (Fig. 5 (G-I)), which is similar to how octahedral rotation narrows the bandwidth of $d_{xy}$ (*23*). This explanation is supported by our density functional theory (DFT) calculation results in Fig. 5I (*26, 27*). There are several energy regions (yellow shaded area in Fig. 5I) in which both $d_{xy}$ and $d_{yz/zx}$ have a peak in DOS. This coincidence is an evidence that $d_{xy}$ (γ-band) and $d_{yz/zx}$ (α-, β-bands) are mixed and hybridized. In other words, octahedral tilting serves as a "scissor" that cuts $t_{2g}$ bands into pieces of narrow bands via hybridization. The resulting narrow bands provides a sufficient condition for the generation of the OSMP.

With the understanding of the band-width reduction mechanism described above, it is important to understand why the β-band is suppressed simultaneously with the OSMP. One may consider the possibility that a Mott-localization may occur in the β-band in a similar fashion to that of the OSMP in the γ-band. However, while the electron occupation number for the γ-band ($n_\gamma \sim 1.5$) is appropriate for Mott localization in a doubled unit cell (*10*) (see section S7), the electron numbers of the α- ($n_\alpha \sim 1.8$) and β- ($n_\beta \sim 0.7$) bands are inappropriate for Mott localization (*9, 10*). Therefore, a mechanism other than Mott localization is needed to explain the suppression of the β-band. We can gain insight into the origin of β-band suppression from previous studies (*4, 6-8, 15, 16*). Some experimental results on 3*d* iron-based superconductors (IBS) suggest that the emergence of OSMP leads to KH between itinerant ($d_{yz}/d_{zx}$) and localized ($d_{xy}$) bands (*4, 15, 16*). Interestingly, CSRO ($0.2 \leq x < 0.5$) exhibits the OSMP (similar to IBS), as well as HF-like behavior (*6-8*); thus, the KH mechanism should be considered.

As seen in Fig. 4 (A and B), the β-band dispersion is significantly renormalized in a similar way to KH, as schematically illustrated in Fig. 4D. Here, the OSMP-driven localized γ-band plays the role of the localized band in KH, and the β-band provides itinerant electrons. Therefore, as the OSMP is strengthened from $x = 0.5$ to $x = 0.2$, the β-band is renormalized to β′ via KH with the γ-band. In this way, as shown in Fig. 2, the OSMP and KH vary simultaneously with changes in $x$ and $T$. Thus, suppression of the β-band can be understood as a result of incoherent-to-coherent crossover due to KH (*28, 29*). Our FS and Fermi momentum data (Figs. 1 and 4) also support the KH scenario. The β-band FS is reduced with OSMP (Fig. 4 (E and F)) and the electron occupancy ($n$) of the β-band decreases from 0.72 ($x = 0.5$) to 0.63 ($x = 0.2$) with OSMP.

Furthermore, our ARPES results reveal that the β- and γ-bands are hybridized at temperatures lower than 40 K (Fig. 3D). Considering the logarithmic temperature dependence of Kondo effects, this temperature (< 40 K) is of a similar order to the incoherent-to-coherent crossover temperature $T^*$ (14 K for $x = 0.2$) for resistivity (*6, 8, 30*) and the antiferromagnetic peak temperature $T_p$ (12 K for $x = 0.2$) for magnetic susceptibility (*8, 30*) (see section S4). Therefore, hybridization between the β- and γ-bands is likely due to KH which usually accompanies both incoherent-to-coherent crossover and antiferromagnetic states. Moreover, the theoretically estimated $T_K$ value is in agreement with our experimentally observed value. HF systems exhibit scaling behavior with respect to $T_K$ which is given as (*5*)

$$\gamma_S \approx \frac{R \log 2}{T_K} \approx \frac{10{,}000}{T_K} [mJ/(K^2 \cdot mol)] \qquad (1)$$

where $\gamma_S$ and $R$ are the Sommerfeld coefficient and gas constant, respectively. The $\gamma_S$ value is about 200-250 $mJ/(K \cdot mol)$ for CSRO ($0.2 \leq x \leq 0.5$) (*7*) and thus the estimated $T_K$ is about 40-50 K, which is consistent with our experimentally estimated value (40 K). In this regard, the β-band suppression is likely due to KH between γ- (localized) and β- (itinerant) bands.

Then, the next question is why only the β-band is involved in KH, while the α-band remains unaffected. This phenomenon can be understood by reference to momentum-dependent-interaction theory, which is essential for explaining ferromagnetic-Kondo systems (*29, 31-34*). In that theory, the proximity of two bands in momentum space is one of the most important factors leading to interactions between them. As can be seen in Fig. 2, the β- and γ- bands are located close in momentum space. Therefore, KH can occur more easily than with the α-band, as illustrated in Fig. 5D.

**Discussion**

CSRO was the first material for which OSMP was proposed, but the very existence of the OSMP in CSRO has not yet been universally agreed on (*10, 11, 19*). The critical reason for the controversy may arise from the fact that the OSMP gap appears as a *soft* gap rather than a hard gap. As can be seen in Fig. 3, $x$- and $T$-variations lead to a gradual spectral weight transfer from a lower to a higher binding energy, rather than in the form of a sudden opening of a hard gap. In other words, suppressed spectral weight as well as remnant quasiparticle peak intensity (*11*) are coincidentally observed in the OSMP of CSRO, explaining both observations of suppressed spectral weight (*10*) and the remnant γ-band at the Fermi level (*11*) (see section S1). Moreover, investigation of the soft gap requires quantitative analysis with a reference point where the gap is closed ($x = 0.5$). Therefore, we speculate that the absence of the reference point ($x = 0.5$) data as well as the use of improper normalization methods (see section S1 for a detailed explanation) may have led previous studies to discrepancies in the interpretations (*10, 11, 19*). Our systematic $x$- and $T$- dependent studies not only settle down this issue by demonstrating the gradual evolution of the OSMP but also provide clues to the microscopic mechanism of the OSMP by demonstrating the coincidence between octahedral tilting and OSMP.

The reason why the OSMP appears as a soft gap can be attributed to the Hund coupling in 4$d$-orbital systems (*35*). The Hund coupling in 4$d$ CSRO is enough to generate orbital-selective behavior but is not strong enough to create a hard gap in the system (*35*). The soft gap nature, which can be viewed as an intermediate state between metal and a Mott insulator with a hard gap, may enable us to observe the $T$-dependent spectral weight transfer (Figs. 2 and 3), which is not typically seen in Mott insulators with a hard gap behavior. There have been many recent attempts to understand the moderate nature (in

Coulomb repulsion and spin orbit coupling) of 4$d$ TMOs which exhibit various exotic phenomena (*36-39*). We believe that results of our study on 4$d$ TMO can provide information on how the Hund coupling plays a role in the determination of the electronic structure, thereby will enable us to understand diverse and exotic phenomena in 4$d$-orbital compounds.

Another important aspect to be discussed is the relevance of the OSMP to the Mott insulating state in Ca$_2$RuO$_4$. Previous theoretical (*40, 41*) and experimental (*42-44*) results in CSRO reveal that the Mott transition in Ca$_2$RuO$_4$ (or $x < 0.2$ of CSRO) is triggered by structural transition (*L-Pbca* to *S-Pbca*). The Mott state exists only in the *S-Pbca* phase, because the *L-* to *S-Pbca* transition leads to a full occupation of $d_{xy}$, which in turn gives the essential condition for the formation the Mott state in the other two bands, *i.e.*, half-filled α- and β-bands. It is interesting to note that a uniaxial strain transforms the Mott state (*S-Pbca*) to a metallic phase (*L-Pbca*) in Ca$_2$RuO$_4$ (*44*) and that the Fermi surface of the metallic phase is remarkably similar to ours in the OSMP phase shown in Fig. 1C. The similarity between the two cases suggests that there may be a correlation between the OSMP and Mott state, which calls for further investigations on this issue.

Finally, our work also has important implications for the aspects of Kondo physics as well. Even though there have been several proposals of heavy mass behavior in CSRO (*6-8*), absence of quantitative and comprehensive electronic structure studies has hindered observation of direct evidence for KH. Our systematic electronic structure studies on CSRO not only reveal the signature of KH in a 4$d$-orbital system for the first time, but also suggest that the OSMP enables the KH via interaction with other itinerant 4$d$-bands. Our work advances understanding of the OSMP and Kondo physics in 4$d$ TMOs, and suggests a key role of octahedral tilting in layered perovskite as a control parameter of physical properties.

## Materials and Methods

### Sample growth and characterization

High quality Ca$_{2-x}$Sr$_x$RuO$_4$ ($x = 0.2, 0.3, 0.4, 0.5, 1.0, 2.0$) were grown using the optical floating zone method. Sample quality and stoichiometry were characterized using a physical property measurement system (PPMS), a magnetic property measurement system (MPMS), scanning electron microscopy with energy dispersive X-ray analysis (SEM-EDX), and X-ray diffractometry.

### ARPES measurements

*T*-dependent ARPES measurements were performed at Seoul National University using a He-I photon source ($h\nu = 21.2\ eV$) and the MERLIN beam line (BL) 4.0.3 of the Advanced Light Source, Lawrence Berkeley National Laboratory using both linearly horizontal (π) and vertical (σ) polarizations of a UV photon source ($h\nu = 70\ eV$). Spectra were acquired using R4000 (SNU) and R8000 (BL 4.0.3) electron analyzers, respectively. Cleaving of the samples was conducted at 10 K in an ultra-high vacuum better than $5 \times 10^{-11}\ Torr$.

### DFT calculations

To obtain the density of states, we performed the first-principles density functional theory calculations using the Perdew-Burke-Ernzerhof functional as implemented in the VASP (*26, 27*). The structural parameters and lattice constants are employed from (*17*).

**Acknowledgments**

**Acknowledgments: Funding:** This work was supported by the Institute for Basic Science in Korea (Grant No. IBS-R009-G2). **Author contributions:** M.K. and W.K. conceived the work. M.K., W.K., Y.K., H.R., J.J. and B.S.K. performed ARPES measurements with the support from J.D.D., and M. K. and D. C. analyzed the data. Samples were grown and characterized by M.S.K., J.Y.K. and D.J.S., with support from Y.Y.. Theoretical studies and density functional theory calculations were performed by C.H.K.. All authors discussed the results. W.K. and C.K. led the project and manuscript preparation with contributions from all authors. Additional information, correspondence and requests for materials should be addressed to W. Kyung (specialtoss@gmail.com) and C. Kim (changyoung@snu.ac.kr). **Competing interests:** We declare that we have no competing interests. **Data and materials availability:** All data needed to evaluate the conclusions in the paper are present in the paper and/or the Supplementary Materials. Additional data related to this paper may be requested from the authors.

**Figures and Tables**

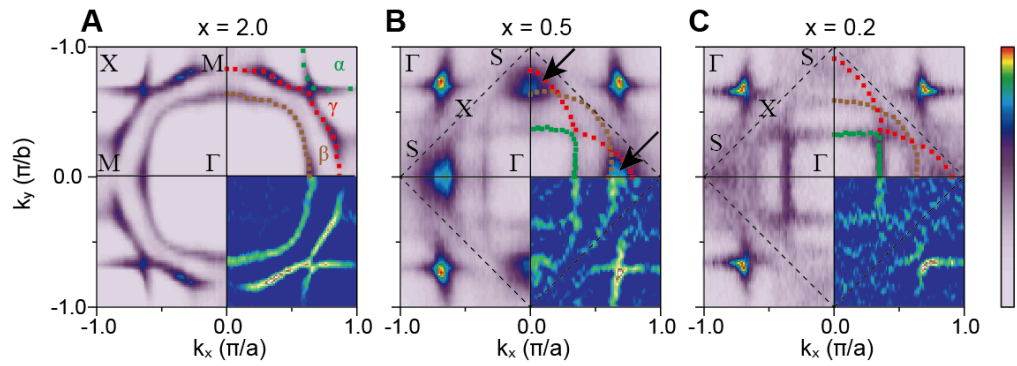

**Fig. 1. Fermi surfaces of $Ca_{2-x}Sr_xRuO_4$.** (A-C) Fermi surfaces (FSs) of $x = 2.0$ (A), 0.5 (B), 0.2 (C) measured using angle-resolved photoemission spectroscopy (ARPES) at $T = 10$ K using $\pi$-polarized light. Color-coded dots indicate the Lorentzian-fitted peak positions of α- (green), β- (brown) and γ- (red) FS pockets obtained from the momentum distribution curves (MDCs). The black dashed lines in (B and C) indicate the reduced Brillouin zone (BZ) induced by octahedral distortions. High-symmetry points in the corresponding symmetry are marked for each doping condition. The right lower part of each panel shows a 2D-curvature plot (*45*) of the FS map.

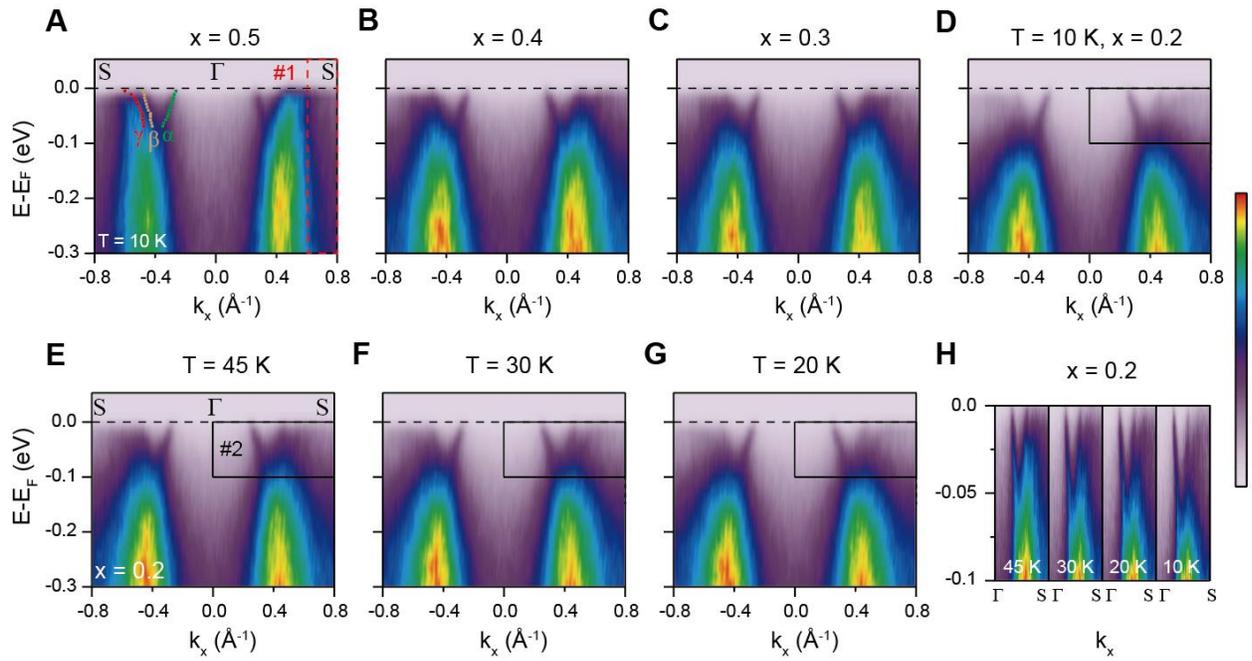

**Fig. 2. Doping- ($x$) and temperature- ($T$) dependent evolution of electronic structure.**
(**A**-**D**) ARPES images along S-Γ-S for $x$ = (A) 0.5, (B) 0.4, (C) 0.3, and (D) 0.2 measured at $T$ = 10 K using π-polarized light. (**D**-**G**) ARPES images of $x$ = 0.2 along S-Γ-S measured at $T$ = (E) 45, (F) 30, (G) 20, and (D) 10 K. (**H**) ARPES images of region #2 (black rectangles) in (D-G); a different color scale is sued to make the $T$-dependence clearer. Note the soft gap opening.

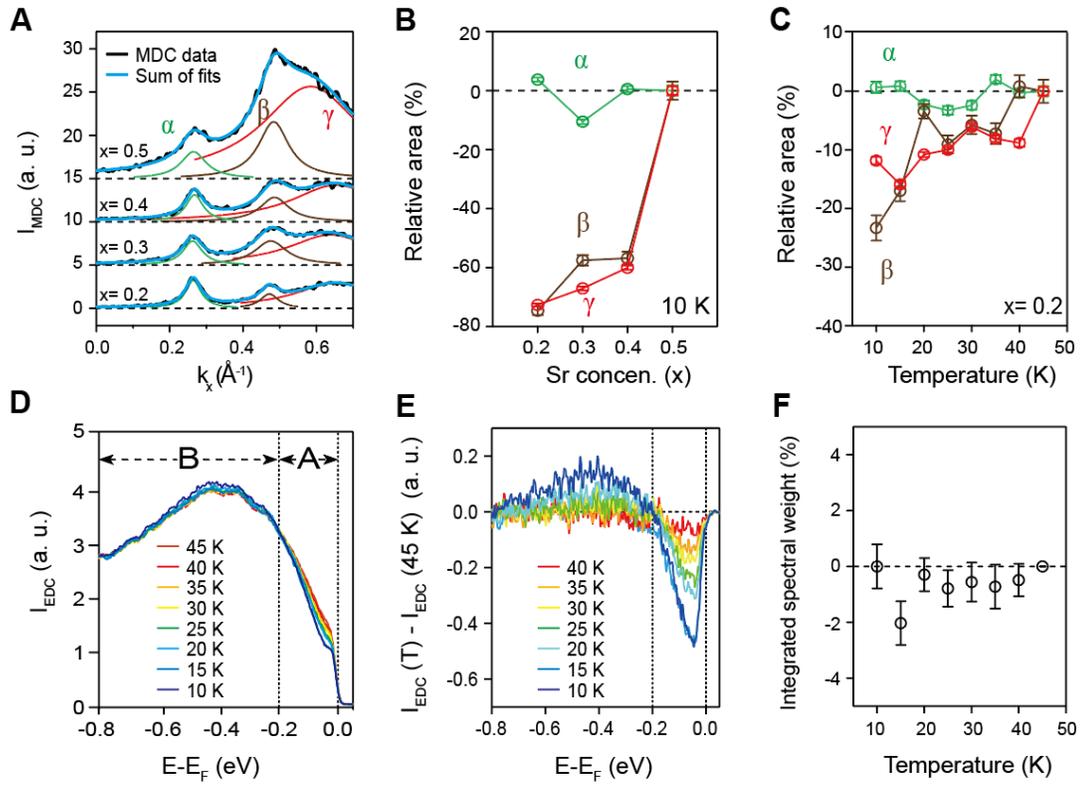

**Fig. 3. Band-selective suppression and spectral weight transfer.** (**A** and **B**) Sr concentration ($x$)-dependent (A) normalized momentum distribution curves (MDCs) at the Fermi level ($E_F \pm 10$ meV) along S-Γ-S. The solid colored lines are the Lorentzian fits (green: α, brown: β, red: γ). (**B**) The $x$-dependent relative change in the Lorentzian fit area of (A) compared to that of $x = 0.5$. (**C**) The $T$-dependent relative change in the Lorentzian fit area derived from the $T$-dependent MDCs (see section S3) compared to that at $T = 45$ K. (**D**) The $T$-dependent integrated energy distribution curve (EDC) of $x = 0.2$ (from region #1 of Fig. 2A), and that (**E**) with the EDC at 45 K subtracted. (**F**) The integrated EDC area (relative to the total integrated area of 45 K in (D)) in the region $-0.8 < E-E_F$ (eV) $< 0.0$ from (E) for each $T$. The zero values at each $T$ suggest spectral weight conservation, implying spectral weight transfer from low- to high-binding energy. A detailed explanation of the normalization method is provided in section S1.

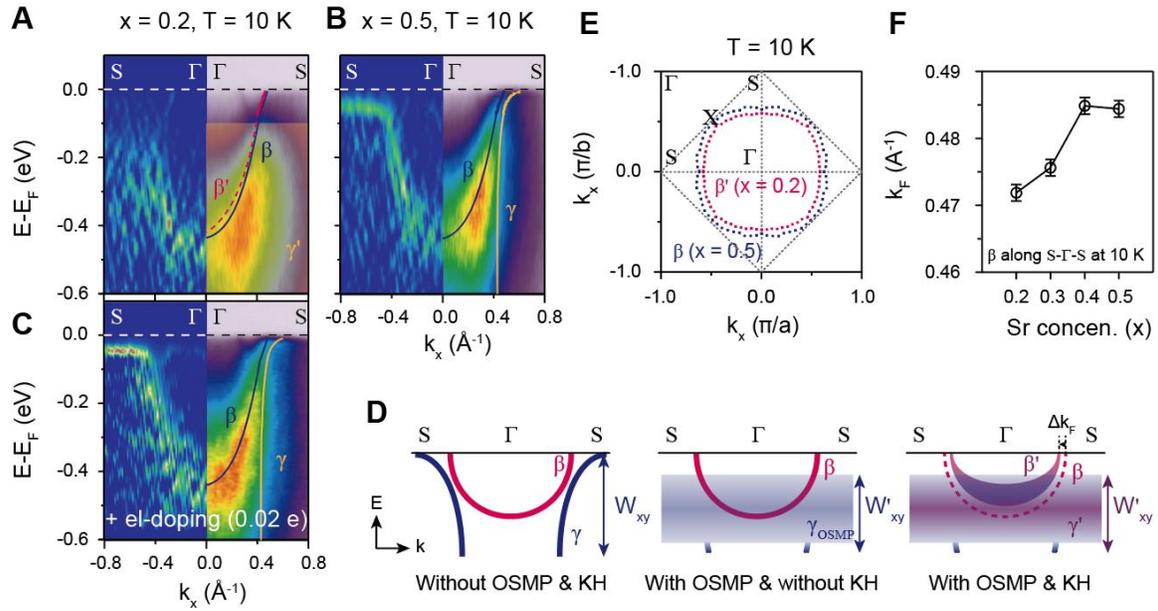

**Fig. 4. Breakdown of the orbital-selective Mott phase (OSMP) with electron doping and illustration of Kondo-hybridization (KH) among the 4d $t_{2g}$ orbitals.** (A-C) ARPES images (right) and corresponding 2D curvature plots (*45*) (left) along S-Γ-S for (A) $x = 0.2$, (B) 0.5, and (C) 0.2 with an electron doping value of 0.02 measured at $T = 10$ K using σ-polarized light. For (C), a monolayer (ML) of alkali metal potassium (K) with a value of 0.2 was deposited *in situ* on the surface of the sample of (A) to achieve an electron doping value of about 0.02 electron to the sample. A detailed explanation of the alkali metal deposition method is provided in section S6. The solid colored lines overlaid on (A-C) are guidelines for the bands (blue: β-band, yellow: γ-band) extracted from Fig. 2; the dashed red line in (A) denotes the β'-band renormalized due to the KH. The yellow shaded area in (A) indicates the Mott localized γ-band (γ'). The γ-band dispersion is taken from Fig. 2; those data were obtained using π-polarized light. (D) Schematic diagrams illustrating the evolution of the β- and γ-bands in three different cases: (left) without the OSMP and KH; (center) with the OSMP and without KH; and (right) with the OSMP and KH. (E) Angle-dependent Fermi momentum ($k_F$ values) without ($x = 0.5$, β) and with ($x = 0.2$, β') the KH extracted from Fig. 1. (F) Doping-dependent $k_F$ of the β-band extracted from Fig. 3A.

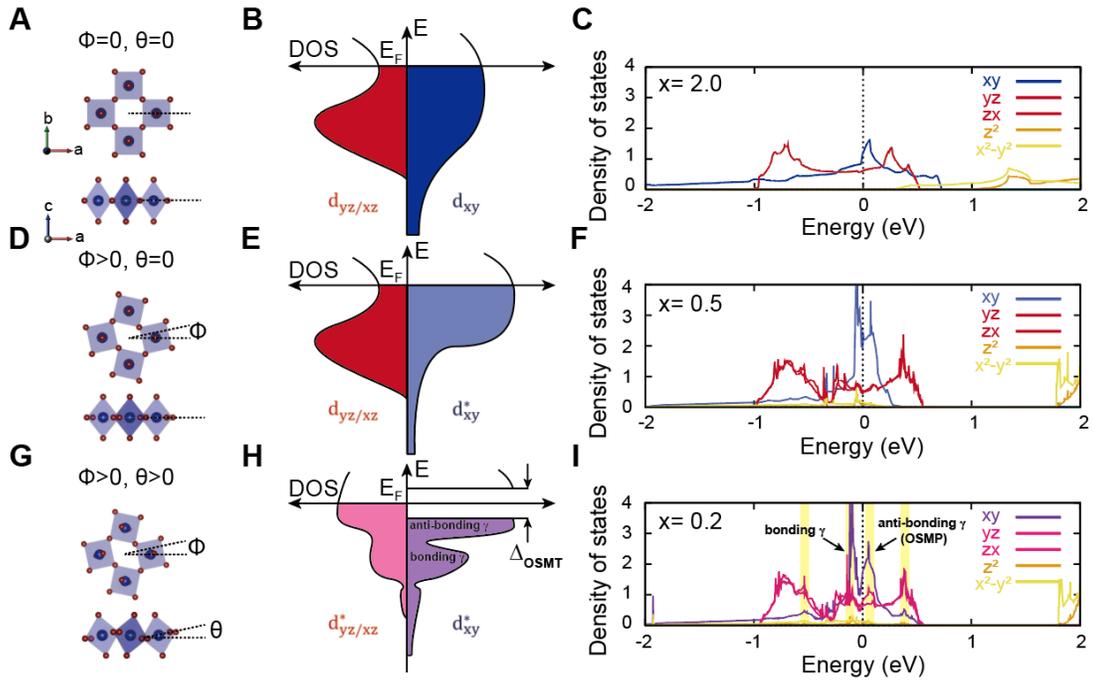

**Fig. 5. Effect of octahedral distortions on the electronic structure of CSRO.** (**A**, **D**, **G**) Top and side view of crystal structures in CSRO (A) without distortion ($1.5 \leq x \leq 2.0$), (D) with rotation ($0.5 \leq x < 1.5$) and (G) with both rotation and tilting ($0.0 \leq x < 0.5$). (**B**, **E**, **H**) Schematic electronic structural changes corresponding to each distortion types: (B), (E), and (H) corresponds to (A), (D), and (G), respectively. (**C**, **F**, **I**) Orbital-dependent ($4d$) density of states (DOS) at (C) $x = 2.0$, (F) $x = 0.5$, and (I) $x = 0.2$ from first-principles density functional theory (DFT) calculations (*26, 27*). Yellow shaded regions in (I) indicate the binding energy regions in which $d_{xy}$ and $d_{yz/zx}$ have a coincident DOS peak at same energy.

# Supplementary Materials for

## Observation of Kondo hybridization with an orbital-selective Mott phase in 4$d$ Ca$_{2-x}$Sr$_x$RuO$_4$

Minsoo Kim, Junyoung Kwon, Choong H. Kim, Daun Chung, Younsik Kim, Hanyoung Ryu, Jongkeun Jung, Beom Seo Kim, Dongjoon Song, Jonathan D. Denlinger, Moonsup Han, Yoshiyuki Yoshida, Takashi Mizokawa, Wonshik Kyung, * Changyoung Kim*

*Corresponding author. Email: specialtoss@gmail.com (W.K.); changyoung@snu.ac.kr (C.K.)

**This PDF file includes:**

## Section S1: Intensity normalization for the ARPES data

To compare the doping ($x$)- and temperature ($T$)- dependent results, the intensity of the angle-resolved photoemission spectroscopy (ARPES) data must be normalized in a systematic way. For an accurate comparison, a normalization area is selected in which bands do not cross. To obtain the normalization factor, we plot momentum-integrated EDCs near the S-point ($0.6 < k_x$ (Å$^{-1}$) $< 0.8$, region #1 in Fig. 2A) as shown in fig. S1A. A binding energy (BE) region indicated in yellow-shaded area between 0.8 and 1.2 eV is selected for normalization since this region has negligible $x$- or $T$-dependence. In brief, all the ARPES data ($x$- and $T$-dependent) are normalized by the intensity of the region ($0.6 < k_x$ (Å$^{-1}$) $< 0.8$, $-1.2 < E - E_F$ (eV) $< -0.8$). Regarding the effect of the normalization method on the results, figure S1 (C and D) show that the $T$-dependent spectrum does not exhibit spectral weight transfer (the OSMP), but does show $T$-dependent breakdown of quasiparticles without OSMP, consistent with previous reports (*11, 19*) (fig. S1D). Therefore, we conclude that the OSMP and QP breakdown may co-exist independently in the γ-band. However, to observe the OSMP, intensity normalization at a binding energy higher than the energy scale of the OSMP is essential.

## Section S2: Doping ($x$)-dependent evolution of orbital-selective Mott phase

The $x$-dependent evolution of the orbital-selective Mott phase (OSMP) in the γ-band is shown in fig. S2. In fig. S2A, the OSMP driven soft gap opens and the spectral weight around $E_F$ is gradually suppressed from $x = 0.5$ to 0.2. The OSMP gap size seems to be around 0.4 eV where the EDC peak increases with spectral weight transfer. To demonstrate the gradual evolution of OSMP, $x$-dependent spectral weight ratios (compared to $x = 0.5$, critical point of OSMP) at $E_F$ are plotted in fig. S2B. Stronger suppression of the spectral weight indicates strengthened OSMP. As shown in fig. S2C ($x$-dependent octahedral rotation / tilting angles (*17, 46*)), the strength of the OSMP (spectral weight suppression of the γ-band) seems to be proportional to the tilting angle.

On the other hand, the EDC at $x = 1.0$ does not show Mott-like (lower to higher BE) spectral weight transfer as well as spectral weight suppression around the $E_F$ (fig. S2A). Since the octahedral tilting angle is zero and the rotation angle varies in the region between $x = 0.5$ and $1.0$ (fig. S2C), we conclude that the OSMP does not occur only with octahedral rotation, hence octahedral tilting is the key to triggering the OSMP.

**Section S3: Temperature ($T$)-dependent evolution of orbital-selective Mott phase**

The $T$-dependent MDCs exhibit orbital-selective spectral weight suppression at $E_F$. As can be seen in fig. S3A, the β- and γ-bands show spectral weight suppression, but the α-band remains almost unchanged. The $T$-dependent orbital selectiveness is more clearly seen in fig. S3B, which shows the MDCs subtracted from that of 45 K.

**Section S4: Electric and magnetic properties of Ca$_{2-x}$Sr$_x$RuO$_4$ ($0.2 \leq x \leq 0.5$)**

To characterize the physical properties of our bulk CSRO crystals, the $T$-dependent in-plane resistivity with the 4-probe method and magnetization with an applied magnetic field along the c-axis were measured using a physical property measurement system (PPMS) and a magnetic property measurement system (MPMS), respectively (fig. S4). Our results (consistent with previous studies (*6-8, 30, 47*)) show characteristic temperatures, the incoherent to coherent crossover temperature ($T^*$, in resistivity) and the antiferromagnetic correlation-driven peak temperature ($T_P$, in magnetic susceptibility). These $T$-dependent slope changes (below/above $T^*$ and $T_p$) in resistivity and magnetic susceptibility are observed in various HF $d$-electron (*3-5*) as well as $f$-electron (*48-52*) systems, related to the hybridization between localized and conduction electrons (*3-5, 48-52*). This strongly suggests a possible Kondo-hybridization (KH) scenario in CSRO.

**Section S5: Breakdown of the OSMP induced by electron doping from alkali metal (K) evaporation**

Alkali metal evaporation is widely used to impart an electron-doping effect on a crystal (*21, 22, 53*). Previous ARPES studies on $Sr_2IrO_4$ (*21, 22*) showed that additional electrons from potassium (K) lead to the sudden breakdown of the Mott-localized state. Motivated by those works (*21, 22*), we performed electronic structure measurements with K evaporation to determine whether the OSMP in CSRO ($x = 0.2$) breaks down in the same way as the Mott state in $Sr_2IrO_4$. As shown in fig. S6, the OSMP soft gap becomes closed with infinitesimal K coverage ($< 0.2$ monolayer, ML), and exhibits recovered dispersive bands. This implies that the origin of spectral weight suppression in the $\gamma$-band is Mott-localization.

**Section S6: $x$-dependent electron occupancies in each band**

Half-filled electron occupancy is essential for Mott localization. In fig. S6E, we extracted the electron occupancies of each band. The $\gamma$-band at $x = 0.5$ has about 1.5 electrons, which satisfies the half-filled condition in the doubled unit cell scenario (*10, 25*), while the other $\alpha$ (1.8) and $\beta$ (0.7) bands are far from the half-filled condition. The electron number of the $\beta$-band decreases from $x = 0.5$ to 0.2. This behavior is likely due to the KH-induced band renormalization.

**Fig. S1.**

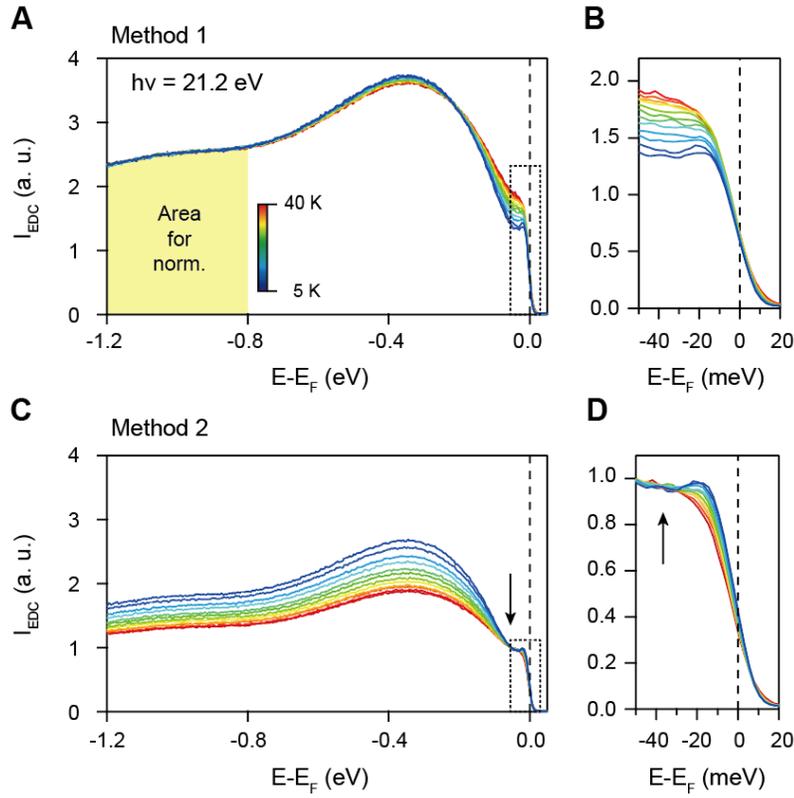

**Fig. S1. *T*-dependent integrated EDCs for *x* = 0.2 with different normalization methods.** (A-D) *T*-dependent integrated energy distribution curves (EDCs) of *x* = 0.2 normalized based on (A) the yellow shaded area (Method 1) in region #1 (0.6 < $k_x$ (Å$^{-1}$) < 0.8) of Fig. 2A and (C) the spectral intensity at E-$E_F$ = - 40 meV (Method 2), as indicated by the black arrow. (B) and (D) are magnified views of the black dotted rectangles in (A) and (C), respectively. We used Method 1 to normalize our data in the main text.

**Fig. S2.**

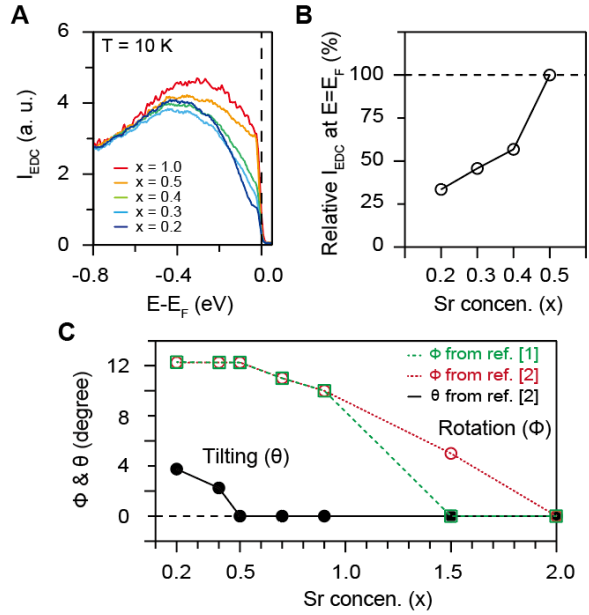

**Fig. S2. Doping (*x*)-dependent evolution of orbital-selective spectral weight suppression.** (**A**) Plot for momentum-integrated EDCs (region #1 in Fig. 2A, $x = 1.0, 0.5, 0.4, 0.3$ and $0.2$) and (**B**) *x*-dependent spectral weight intensity ratio (compared to $x = 0.5$) at the Fermi energy ($E_F$) extracted from (A). (**C**) Plot of *x*-dependent $RuO_2$ octahedral rotation ($\Phi$, open square (*17*) and circle (*46*)) and tilting ($\Theta$, solid circle (*46*)) angles obtained from (*17, 46*).

**Fig. S3.**

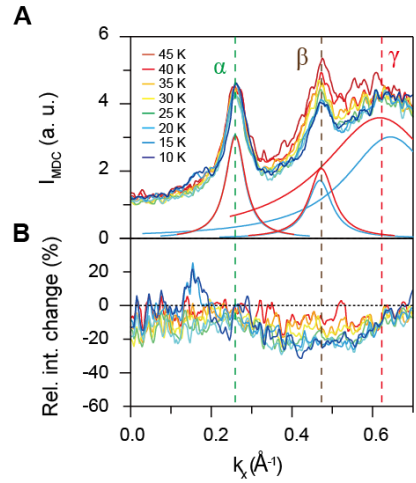

**Fig. S3. *T*-dependent evolution of orbital-selective spectral weight suppression.** (**A**) *T*-dependent normalized momentum distribution curves (MDCs) of $x = 0.2$ at $E_F$ (±10 meV) along Γ-S. The colored solid lines are fitted Lorentzian functions of the α-, β- and γ-bands from MDCs measured at 15 K (blue) and 45 K (red). The colored dashed lines indicate the $k_F$ position of each band. (**B**) *T*-dependent relative intensity changes in MDCs compared to 45 K.

**Fig. S4.**

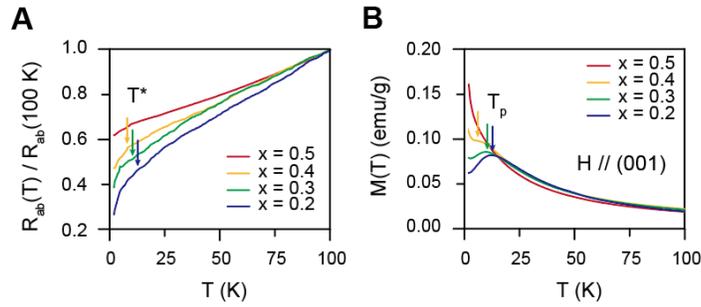

**Fig. S4. In-plane resistivity and magnetic property data of $Ca_{2-x}Sr_xRuO_4$ (CSRO) ($0.2 \leq x \leq 0.5$).** Measured *T*-dependent plots for (**A**) in-plane resistivity normalized at *T* = 100 K and (**B**) magnetization measured with an applied magnetic field (0.1 Tesla) along the c-axis (001). The colored arrows indicate the reported values of *T\** (incoherent to coherent crossover temperature) and $T_P$ (peak temperature in magnetic susceptibility), respectively, obtained from references (*6-8, 30, 47*).

**Fig. S5.**

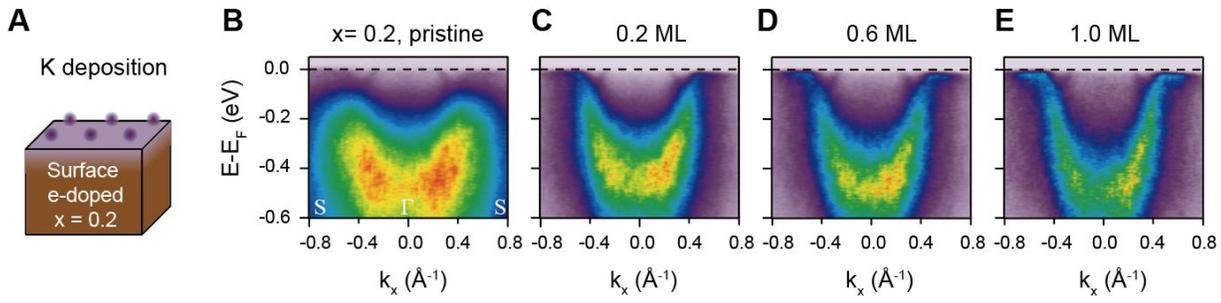

**Fig. S5. Breakdown of the OSMP on K deposition.** (**A**) Schematic illustration of alkali metal (K) deposition on the surface of a Ca$_{2-x}$Sr$_x$RuO$_4$ ($x = 0.2$) single crystal. (**B-E**) ARPES images along S-Γ-S of $x = 0.2$ measured with σ-polarized light with K of (B) 0, (C) 0.2, (D) 0.6 and (E) 1.0 monolayer (ML) coverage.

**Fig. S6.**

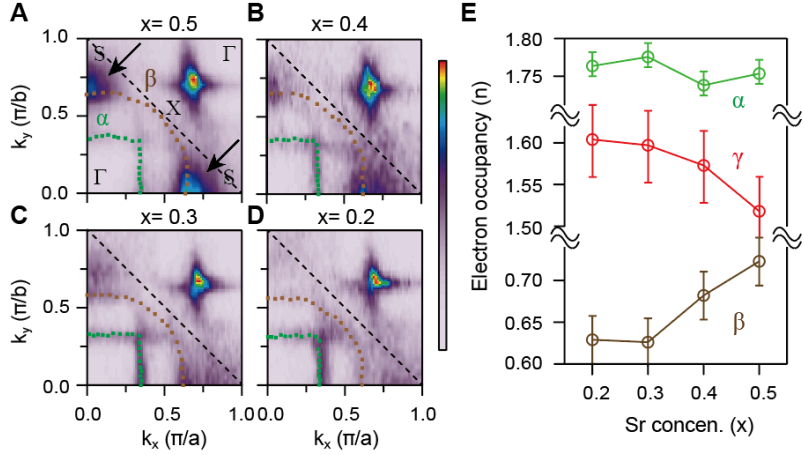

**Fig. S6. *x*-dependent Fermi surfaces and electron occupancies.** (**A-D**) Fermi surfaces (FSs) of (A) $x = 0.5$, (B) 0.4, (C) 0.3, and (D) 0.2. As $x$ decreases from 0.5 to 0.2, the $\gamma$-band intensity (arrow) weakens and eventually disappears (OSMP). Colored dots indicate FS contours of the α- (green) and β- (brown) bands obtained from a Lorentzian fit of MDCs. Black dashed lines indicate the reduced Brillouin zone (BZ) due to octahedral distortions. (**E**) The electron occupancies of each band are plotted as a function of $x$ extracted from Luttinger's theorem (*54*). Since the $\gamma$-band occupancies cannot be directly obtained due to the OSMP, they are calculated by subtracting the α- and β-band occupancies from 4.